\documentclass[12pt]{iopart}

\usepackage{amssymb}

\def\beq{\begin{eqnarray}}
\def\eeq{\end{eqnarray}}
\def\ln{\,\mbox{ln}\,}

\def\be{\beta}
\def\ch{\chi}

\def\si{\sigma}

\def\La{\Lambda}

\begin{document}

\title[Cosmological Constant Problems and Renormalization Group]
{Cosmological Constant Problems and Renormalization Group}

\author{Ilya L. Shapiro $^{a}$ \footnote{
Also at Tomsk State Pedagogical University, Tomsk, Russia; \
electronic address: shapiro@fisica.ufjf.br},
 \  Joan Sol\`a $^{b}$ \footnote{
Electronic address: sola@ifae.es}}

\address{(a) Departamento de F\'{\i}sica -- ICE,
Universidade Federal de Juiz de Fora, MG, Brazil

(b) Dep. E.C.M.  Universitat de Barcelona, and C.E.R. for
Astrophysics, Particle Physics and Cosmology,
\thanks{Associated
with Instituto de Ciencias del Espacio-CSIC.}, Diagonal 647,
Barcelona, Catalonia, Spain}


\begin{abstract}
The Cosmological Constant Problem emerges
when Quantum
Field Theory is applied to the gravitational theory, due to the
enormous magnitude of the induced energy of the vacuum. The unique
known solution of this problem involves an extremely precise
fine-tuning of the vacuum counterpart. We review a few of the
existing approaches to this problem based on the account of the
quantum (loop) effects and pay special attention to the ones
involving the renormalization group.
\end{abstract}


\section{Introduction}
 There are two main reasons
 for introducing the cosmological constant,
 one theoretical and one experimental. The theoretical reason is that
 there are many distinct sources of the cosmological constant (CC) in
 Quantum Field Theory
 (QFT) and Particle Physics.
 Below we shall discuss some of them. The experimental/observational evidence
 of an accelerated expansion of the universe comes from the
 type-Ia supernova observations\,\cite{Supernovae}, from the CMB data\,\cite{WMAP3Y}
 and also from the  available rich information about the galaxy
 distribution\,\cite{LSS}.
 Usually the nonzero vacuum energy is referred to as Dark
 Energy (DE), because it does interact with the matter
 content of the universe only gravitationally and its ultimate
 nature is unknown. There are many
 candidates to play the role of DE (e.g. quintessence or phantom
 energy\,\cite{quintessence,Phantom}), but due to the
 mentioned theoretical arguments, the CC is the main candidate
 for the role of DE. However the situation is spoiled by the CC
 problems, including the problem of fine-tuning \cite{weinberg89}
 and the related coincidence problem\,\cite{Coincidence}. One can easily identify
 these problems as a hierarchy problems arising due to the
 enormous scale difference between large scale gravitational
 physics (cosmology) and the short scale characteristic of
 high energy particle physics -- see \cite{nova} for a detailed
 discussion.
 In this situation even very small quantum corrections to
 the CC may be relevant and, in principle, lead to a certain
 observable consequences. Indeed, the renormalization group (RG)
 is the most economic way to parametrize these quantum
 corrections and investigate their phenomenological impact.
 The purpose of this short review is to put together a few
 different approaches to the CC problem based on the
 renormalization group. The review is mainly based on the
 original papers \cite{nova,lam,cosm,CCfit,Gruni}.

\section{Why do we need CC in QFT?}

There are two sources of the vacuum energy which have essentially
distinct origin. One of them is the classical gravitational
action. Besides the Einstein-Hilbert term, this action can include
additional terms, both local and non-local. The simplest of them
is the CC. If we consider the quantum theory of matter fields on
the classical gravitational background, the consistency
requirement is that the action of vacuum has the form (see, e.g.
\cite{birdav,book}) $\, S_{vac} = S_{EH} + S_{HD}$, where 
\beq
S_{EH} = - \int d^4x\sqrt{-g}\,\left\{\, \frac{1}{16\pi
G_{vac}}\,R + \La_{vac} \,\right\}\,, \label{21} 
\eeq 
and
$\,S_{HD}\,$ include fourth derivative terms\footnote{We denote by
$\Lambda$ the CC density itself, sometimes indicated as $\rho_V$ 
in the literature.}. Without any of the mentioned terms (in
particular the $\Lambda$ term) the theory is not renormalizable
\,\cite{Stelle77}. When looking at the typical cosmic scale energy, 
it seems that the UV divergences are irrelevant. The cosmic scale 
energy can be characterized by the Hubble parameter 
$H_0 \sim 10^{-42} GeV$, which is about 30 orders of magnitude 
smaller that the mass $m_\nu$ of the lightest
neutrino or that the energy of the relic radiation photon.
However, the action of vacuum should be the same at all periods of
the history of the universe, including the very early epoch when
the Hubble parameter was much greater and UV divergences become a
serious problem if the theory would be non-renormalizable.

The cosmological term is thus an unavoidable element of the vacuum
action (\ref{21}), meaning that if the DE is modeled in
alternative ways (e.g. through quintessence) the contribution from
the CC term is always there and must necessarily be taken into
account\,\cite{nova}. The natural question is what is the natural
magnitude of the vacuum CC. In order to address this problem, let
us remember the renormalization group equation for the CC \beq
\mu\,\frac{d\La_{vac}}{d\mu}\,=\,\be_\La
\,=\,\frac{m_s^4}{2}\,-\,2m_f^4\,, \label{RG CC} \eeq where we
just took into account the contributions of a massive scalar and a
fermion. $\,\mu\,$ is the typical energy of the external field
(graviton, in the case). The last expression shows that the
natural range of $\La_{vac}$ is given by the fourth power of the
mass of the heaviest particle. Obviously this will produce a
serious conflict with the measured value of the CC, unless the
only contribution would be from neutrinos of $m_{\nu}\sim
10^{-3}\,eV$\,\cite{cosm}. In order to reduce this estimate we
need a cancellation between bosons and fermions in Eq.\, (\ref{RG
CC}). Supersymmetry (SUSY)\,\cite{SUSY}  may, therefore, help to
reduce the minimal admissible value of $\La_{vac}$. However, at
low energies SUSY is known to be broken, because the proper model
for the physics up to the Fermi scale is the Minimal Standard
Model (SM) of particle physics. Hence the RG equation (\ref{RG
CC}) can only be applied for the energies comparable to the
typical scale of the SM and, in order to be compatible with the
high energy running, the magnitude of the vacuum CC density should
be of the order of the fourth power of the electroweak scale,
namely $\sim 10^8 GeV^4$.

Another source of the CC is the induced action of gravity
\cite{zeld}, e.g. the one which emerges through the electroweak
(EW) Spontaneous Symmetry Breaking (SSB). In the ground state of
the Higgs potential of the SM, $\,V(\phi)\,=\, - (m^{2}/2)\phi^{2}
+ (f/8)\,\phi^4\,$, the induced CC is
\beq \La_{ind}=<V> = -
\frac{m^4}{2f}=-\frac18\,M_H^2\,v^2\approx - 10^{8}\,GeV^{4}\,,
\label{induced CC} \eeq
where $M_H$ is the value of the Higgs
boson mass and $v\simeq 250\,GeV$ the VEV of the Higgs potential.
Since we know from LEP experiments that $M_H>114 \,GeV$ we obtain
the estimate (\ref{induced CC}). It is remarkable that this
estimate coincides with the natural value of the vacuum parameter
$\La_{vac}$. However, what we really observe is the sum 
\beq
\La_{obs} = \La_{vac} + \La_{ind}\,, \label{obs} 
\eeq 
where the induced quantity $\,\La_{ind}\,$ receives many different
contributions (e.g. from the electroweak SSB, chiral symmetry
breaking and other possible phase transitions, plus quantum
corrections to these contributions) and can not be, in principle,
calculated exactly.

The real problem is that the cancellation of the two
independent contributions $\La_{vac}$ and $\La_{ind}$ (even though
they can be expected to be of the same order of magnitude, as we
have seen above for the EW case) must be extremely precise. By
virtue of the recent astronomical observations the observed value
of the CC density is 
\beq 
\La_{obs} \approx 0.7\,\rho_c \sim
10^{-47}\,{\rm GeV}^4\,. 
\label{obs CC} 
\eeq 
Looking at the
situation from the QFT viewpoint, the cancellation between
$\La_{ind}$ and $\La_{vac}$ can be provided by imposing a
restriction on the independent parameter $\La_{vac}$. This
restriction is nothing but the renormalization condition and it
should be implemented at the cosmic scale $\mu_c$ where the
observations are performed. 
\beq 
\La_{vac}(\mu_c)=\La_{obs} - \La_{ind}(\mu_c)\,. 
\label{8} 
\eeq 
In the last relation the
second term on the \textit{r.h.s.} ($ \La_{ind}(\mu_c)$) is 55
orders of magnitude greater than the observable term. Therefore
the renormalization condition for $\La_{vac}$ at the scale
$\mu=\mu_c$ reads as follows: the quantity $\La_{vac}(\mu_c)$
must be equal to $-\La_{ind}(\mu_c)$ up to the 55th digit, for
otherwise we should never meet the small $\La_{obs}$ which is
actually observed. If we assume that there was another phase
SSB-based transition at a typical GUT scale ($M_X\sim
10^{16}\,GeV$), there will be more than 110 orders of magnitude
difference and, finally, the phase transition at the Planck scale
would give 123 orders. The (extremely huge) problem of {\it why}
the two terms cancel so accurately is the famous CC problem, the
biggest conundrum ever (see \cite{weinberg89} for a classical
review).

\section{CC problem is a hierarchy problem}

The simplest option is just accepting this cancellation as a
fact. One can compare the situation with the great success of the
SM, which has plenty of phenomenological parameters like masses
of the particles etc. Of course, we can not measure any of them
with the 55-order precision, but this only shows that our
``measurement'' of the CC is extremely precise. The origin of
this ``precision'' is nothing but the difference between the MS
Fermi scale where we evaluate the induced CC, and the tiny cosmic
scale where we observe the total CC. Hence, {\it the CC problem
is a hierarchy problem}, which results from the conflict between
the physics at different scales.

In fact, the CC case is even more complicated. Remember
that the Universe is not static and that
the temperature of the relic radiation was much higher in
the past than it is now. In the Early
Universe, there was an epoch when this temperature was
about the Fermi scale $T\propto M_F$. At that temperature,
according to the standard viewpoint,
the symmetry in the potential gets restored
and the induced CC disappears or just becomes many orders
smaller. Thus, in the earlier Universe the
overall CC includes the vacuum contribution $\La_{vac}$
only. This means that our Universe had been created
from the very beginning such that $\La_{vac}$, after the
symmetry breaking phase transition, should cancel the
induced CC with the tremendous $55$-order precision. This
does not look natural at all, and we really have to worry
about this.

In addition to the rapid changes of the CC due to the phase
transition, there should be a renormalization group running of
both induced and vacuum counterparts of the CC. This issue was
discussed in detail in \cite{nova}, see this reference for
details. Here we just notice that the running involves, in one way
or another, the masses of the constituents of the SM. It might
happen that the contribution of some particle is suppressed
because it shows up only at higher loops, but in any case all
physical SM effects occur at scales whose ratio with the Fermi
scale is far away from those 55 orders of magnitude associated to
the CC problem. At this point we can conclude that the CC problem
is something fundamental and that its solution should perhaps
involve also the explanation of the particles mass spectrum.

Below we shall give a brief review of some methods of
solving the CC Problem. There are many reviews (see e.g.
\cite{CCRev}), so we will refer the reader to this
work for the list of existing approaches and mainly
concentrate here on the ones which do not involve higher
dimensions/branes and are related to the quantum effects.

\section{Supersymmetry, strings and anthropic approach}

There were many attempts to solve the CC problems introducing more
symmetries, e.g. supersymmetry (SUSY), which simply forbids the
contributions to the CC because the SUSY vacuum must have zero
energy (see the discussion in \cite{wittenDM} and \cite{weinDM}
on the subject). Unfortunately, from the general perspective SUSY
does not look helpful in solving the CC problem. The reason is
that SUSY is explicitly broken at the electroweak scale, and
therefore it cannot prevent the vacuum energy from getting
contributions similar to the SM ones, see (\ref{induced CC}). To
put in another way: SUSY is a high-energy phenomenon. As a result
we find ourselves in the following situation: while at low
energies SUSY is broken and this leaves no much hope to apply it
for the solution of the CC problem, at high energies SUSY may
effectively apply, but then it cannot solve the CC problem because
this problem possibly does not exist anymore there. Indeed, at
scales near $M_P$ or above the full SUSY $\beta$-function of the
$\Lambda$ term (i.e. the full structure on the \textit{r.h.s.} of
Eq.\,(\ref{RG CC}) at very high energies) is probably zero. Since
there are no induced contributions at those energies, but only
the vacuum term $\Lambda_{vac}$, there are no fine tunings, and
$\Lambda_{vac}$ can naturally be of order $M_P^4$ and remain
peacefully so around those energies simply because it does not run
while $\beta_{\Lambda}=0$.

Another hope is the (super)string theory. However, the choice of
vacuum for the string theory is not unique, at least at the
present day state of knowledge. It might happen that the ``right''
vacuum gives a ``correct'' value of the CC \cite{bousso}. At first 
sight it is unclear how this can affect the low-energy physics, for 
at low energies, we have a very strong experimental confirmations 
that the appropriate theory is the SM and not the string theory. Be 
as it may, at the moment moving from QFT to string theory does not
seem to help much, for after the process of compactification from
$11$ dimensions down to $3$ we are left with a vastly complex
``landscape'' consisting of some $10^{1000}$ metastable
(non-supersymmetric) vacua  where to entertain our choice of the
ground state\,\cite{landscape1,landscape2}. There is an important
new aspect, however. If we consider the existence of a very
complicated vacuum structure of string theory together with the
anthropic hypothesis, there is a hope to arrive at some
consistent picture and maybe even learn some lessons about
fundamental physics.

The anthropic approach is using the ``experimental'' fact of our 
own existence. One can say it produced the main success concerning 
the CC problem. The evolution of density perturbations depends on 
the equation of state of the matter and vacuum content of the 
universe. The formation of galaxies  and stellar systems should 
perform in such a way that friendly conditions for life take 
place. This imposes rigid constraints on the evolution of density 
perturbations in the universe. This evolution depend, along with 
other factors, on the vacuum energy density and hence imposes 
certain constraints on its possible values. In this way, Weinberg 
predicted in 1987 the positiveness of the CC \cite{weinberg87}. 
More sophisticated considerations \cite{Vilenkin} indicated 
greater probabilities of the CC being quite close to the 
astronomically observed value. Taken on 
its own the anthropic approach can not answer the fundamental 
question of {\it why} we should exist at all. But from the 
landscape perspective the answer may be just an existence of 
many universes (a ``multiverse''), most of which had (or have) 
no chance to evaluate or are not visible for us. Let us notice, however, strong divergeneces between different definitions of 
weight functions \cite{antrop1,antrop2,Vilenkin}. In particular, 
they may be relevant for calculating anthropic probabilities 
in the universes with different values of CC.

\section{Auto-relaxation or adjusting mechanisms}

The measurement that the cosmological constant is non-vanishing is
relatively recent, it is only from 1998. Prior to this date the
general belief from the theoretical physics and cosmologist
community was that the cosmological constant had to be exactly
zero. Therefore, it was quite natural to seek for an efficient
adjustment mechanisms in which the value of the vacuum energy
relaxes to zero in a dynamical way. The prototype mechanism to
achieve this dynamical adjustment was to use some scalar field
under some suitable potential or motivated by some symmetry
requirement (e.g. dilatation symmetry). There were a number of
very interesting attempts to create a sort of automatic mechanism
for relaxing the CC. The general idea is to consider a modified
gravity theory, where the effect of cosmological constant is
reduced. Typically the original auto-relaxation models
\cite{Dolgov,PSW,Others} involved a scalar field which moves to
the minima of its potential along with the evolution of the
universe. As an example, the ``cosmon'' field introduced in
\cite{PSW} aimed at a Peccei-Quinn-like adjustment mechanism
based on a dynamical selection of the vacuum state at zero VEV of
the potential, $<V>=0$. More recently these ideas have been
exploited profusely in various forms, such as the so-called
``quintessence'' scalar fields and the like\,\cite{quintessence},
``phantom'' fields\,\cite{Phantom} etc, including some recently
resurrected old ideas on adjusting mechanisms \cite{Barr}. 
The main aim of these dynamical mechanisms is that the induced CC
goes to zero automatically and the problem of a fine-tuning
between the induced and vacuum counterparts may be smoothed or
banished. Of course, with the advent of the high precision
cosmology experiments the hard job of the quintessence community
is to understand why the relaxation point of the quintessence
field is not precisely zero but some extremely small value of the
potential! This value amounts to introduce a small mass for the
quintessence field $\ch$ of order $m_{\chi}\sim H_0\sim
10^{-42}\,GeV$, which is some $17$ orders of magnitude smaller
than the upper bound on the photon mass from terrestrial
experiments! Clearly, this shows the highly artificial character
of the quintessence field from the Particle Physics standards. In
the classical review by Weinberg\,\cite{weinberg89} many of these
dynamical approaches, and their presently insurmountable
difficulties, are discussed in quite some detail. They all end up
with some more or less obvious form of fine-tuning.

\section{Renormalization group and CC Problem}

The renormalization group (RG) is a conventional theoretical tool
for investigating the scale dependence. As we have seen above,
the CC problem is a violent conflict between two scales: the high
energy scale $M$ (typically $M\sim M_P$) where $\La_{vac}$ and
$\La_{ind}$ are defined, and the (low energy) cosmic scale
$\mu_c$ ($\ll M$) where we can observe their sum $\La_{obs}$.
Therefore it is quite natural to consider the CC problem from the
RG perspective. Let us notice that the idea of the renormalization
group solution became quite popular in the last years
\cite{TV,Polyakov}. The first practical realization has been
suggested by Antoniadis and Mottola \cite{antmot} in the
framework of the quantum theory of conformal factor
\cite{odsh91}, which is a direct $4d$ analog of the $2d$ Polyakov
theory. Other realizations are based on different versions of the
IR quantum gravity \cite{TW,ReuterCC,BFLWard}.

An alternative and perhaps the most simple way to achieve
the IR screening of the CC has been suggested in \cite{lam}.
This approach relies on the IR quantum effects of matter
field rather than on the IR quantum gravity effects.
In this section we shall review the proposal of \cite{lam},
trying to reformulate it in a slightly physical way. In
the next section we shall consider even more physical
(and less ambitious) approach to the CC Problem based
on taking the quantum effects of matter fields into account.

Let us suppose that:
\quad
{\it i)} \ the symmetry restoration at $T \propto M_F$
does not happen. Indeed, this means that the Higgs sector
of the SM must be extended such that the non-restoration
becomes possible. Then we live not in the SM vacuum but
in the GUT vacuum.
\quad
{\it ii)} \ The origin of all masses of the fields in the
SM and beyond is the Coleman-Weinberg mechanism in some
GUT model which describes the physics at a very high energy
scale. The quantum symmetry breaking in the field $\Phi$
leads to the induced cosmological $\La$ and inverse Newton
$1/G$ constants. Since all fields are massless, we do
not need to introduce vacuum CC and $G$ and therefore
no need to distinguish $\La_{obs}$ and $\La_{ind}$.
The classical potential for the field
$\Phi$ includes the nonminimal interaction term, since it
is necessary for renormalizability
$V = - \frac{\xi}{2}\,R\,{\Phi}^2+ f {\Phi}^4$.
The induced gravitational quantities are
\beq
(16\pi G)^{-1} \,\,\sim\,\,
< \xi (t) \Phi_0^2 (t) >
\,,\qquad
\Lambda\,\, \sim\,\, - < f(t)\Phi_0^4 (t) >\,.
\label{111}
\eeq
In the last expressions we have introduced the dependence
on the RG parameter $t=\ln(\mu/\mu_0)$.
The explicit form of this dependence is a function of the
GUT model under consideration.
\quad
{\it iii)} \ The fundamental theory
has a huge number of copies ${\cal N}$ of all or some of its
constituents. These fields couple to the scalar $\Phi$
in such a way that the $\,\be$-function for the nonminimal
parameter $\xi$ has the form
$\,\be_\xi = \Big(\xi-\frac16\Big)\cdot {\tilde A}(g^2)$,
where $\,{\tilde A}(g^2)=-A{\cal N}g^2\,$
is a linear combination of the (gauge)$^2$, (Yukawa)$^2$ and
scalar couplings. For simplicity we assume that
the square of Yukawa couplings and scalar couplings
are proportional to the square of the
gauge coupling $g^2$.

We require that $A$ is positive such that
the conformal fixed point $\xi=1/6$ is stable in UV
and the point $\xi=\infty$ to be an attractor in the
IR low-energy limit (see general discussion and references 
to original papers in \cite{book}
and classification of the $SU$(N) and $O$(N) gauge models
in \cite{buya}).
\ {\it iv)} \ For the sake of simplicity, we shall suppose
that the
gauge theory is finite - that is the $\be$-functions for
all coupling constants in the matter fields sector are
zeros. This is not a necessary requirement, and one can
consider another type of theories. For example, similar
model has been recently discussed in
\cite{Jackiw}, based on the quantum theory of scalar field
with the coupling constant growing on in the IR limit.

In the leading log. approximation for $V_{eff}$ one can
use the RG improved classical potential $V$.
In this way we find
\beq
G^{-1} \propto < \xi (t) \Phi_0^2 (t) >
\quad \mbox{and}\quad
\frac{\Lambda}{G} \propto -< f(t)\Phi_0^4 (t) >\,.
\label{16}
\eeq
Due to the gauge dependence, the RG for
$\Phi(t)$ is not sufficient to determine asymptotic of this
effective charge. At the same time we can easily find it
using physical considerations.
In fact, since $G(- \infty)$ have the finite classical value
and \  $\xi(t) \sim exp(- Ag^2t)$ we find that
 $\Phi(t) \sim exp(Ag^2t)$ .
Now we take into account (\ref{16}). Since in the finite
models $\lambda (t) \equiv {\lambda}_0 = const$, we find that
in the IR limit $t \rightarrow -\infty$
\beq
\Lambda \propto < f(t) \Phi_0^4 (t) >
\sim \exp(-2A{\cal N}g^2t)
\quad \mbox{and}\quad
\frac{\La_{IR}}{\La_{UV}}
= \Big( \frac {\mu_{IR}}{\mu_{UV}} \Big)^{-2A{\cal N}g^2}\,,
\label{17}
\eeq
where we assume CC running between the energy scales
${\mu }_{UV}$ and ${\mu}_{IR}$

The result for the $\La_{IR}$ depends on the choice of the
model (that defines $A$) as well as on the region of it's
application. Let us consider the running from the Planck
scale $\,{\mu}_{UV} = M_P \approx 10^{19}GeV\,$ down
to the present-day cosmic scale
$\,{\mu}_{IR} = \mu_c \approx 10^{-42}\,GeV$.
The values of $A$ have been calculated in a number of papers
(see, for example, \cite{book,buya}), and the typical values for
$A$ are between
$\,{1}/{(4\pi)^2}\,$ and $\,{50}/{(4\pi)^2}.$
For the maximal value we obtain
\beq
{\La_{IR}}/{\La_{UV}}\,\,\, \approx\,\,\,
10^{\,-\, 6000 \,{\cal N}\,g^2/(4\pi)^2} \,.
\label{19}
\eeq
Taking $g\approx 10^{-1}$ and ${\cal N}=1$, we can see that
the value of $\Lambda$ is decreasing on about four orders.
But, if we take ${\cal N}=30$ copies of the fields, we arrive
at the tremendous 120 orders, which can solve the CC
problem. The effect would be seen in the astronomic
observations as a very slow decrease of the observable CC
during the last few billion years.

The above model of IR screening for CC does not look like
a natural solution of the CC problem. The need for the great
number of copies of the fields (they must be massless, for
otherwise they just decouple long before the IR limit),
symmetry non-restoration at $T=M_F$ and all masses of the
SM being the result of the dimensional transmutation
at the GUT scale is not very appealing from the
phenomenological point of view. The advantage of the model
is that it is really free of a usual CC fine-tuning.

\section{Decoupling and cosmological constant}

$\,$\quad In this section we shall consider the less ambitious
program than in the previous ones. Namely, we will not try
solving the great CC problem \cite{weinberg89}, neither the
coincidence problem (see, however, the generalized RG model with
cosmon field \cite{LXCDM}). Instead we accept a purely
phenomenological point of view and assume that the particle
physics can be described by the known SM or some its conventional
extension. The question is whether it is possible that even in
this situation the renormalization group may be relevant, that is
whether the observable value of the CC can depend on time due to
the quantum effects. At the first glance this question looks as
something absurd. Usually the low-energy effects of massive
quantum fields manifest the quadratic decoupling at low energies
\cite{AC}. Is it true that, despite the huge difference in the
magnitude of the energy scales the quantum effects can be
relevant for the CC? Curiously, the answer is yes. In this
section we shall present the basic ideas of the approach of
\cite{nova,CCfit} based on the notion of ``soft decoupling'' of
massive fields at low energies in the gravitational sector (see
also \cite{babic}). One can find many important details in these
papers and also in the last related developments in
\cite{Gruni,LXCDM,SS}, in particular the implications of a running
CC on the matter power spectrum\,\cite{CCwave}.

Despite the calculations of decoupling for massive fields on
curved background \cite{apco} were not successful in the CC
sector, but the decoupling in other sectors of the vacuum action
is of the standard quadratic form. Let us use a phenomenological
approach and assume that the quadratic decoupling holds for a CC.
In the present-day Universe one can associate $\mu \equiv H$
\cite{nova}.

Remember from Eq.\,(\ref{RG CC}) that $\be_\La \sim m^4$, $m$
being the mass of a contributing quantum field. Then the
quadratically suppressed expression is \cite{nova,CCfit} 
\beq
H\,\frac{d\Lambda}{dH}\,=\, \beta_{\Lambda}\,=\,
\sum\limits_i\,c_i\,\frac{H^2}{m_i^2}\,\times m_i^4
\,=\,\frac{\si}{(4\pi)^2}\,M^2\,H^2\,, \label{IR CC} 
\eeq 
where $M\,$ is an unknown mass parameter and $\,\si=\pm 1\,$ 
depending
on whether fermions or bosons dominate in the particle spectrum.
Assuming $\,M^2=M_P^2$, we find $\,|\beta_{\Lambda}|\sim
10^{-47}\,GeV^4$, which is close to the existing supernovae and
CMB data for the vacuum energy density. Therefore, the
renormalization group may, in principle, explain a smooth
variation of the vacuum energy without introducing special
entities like quintessence.

Two cosmological models with running CC have been developed in
\cite{CCfit,Gruni}, based on the original RG framework of
\cite{nova}. Further developments around these models have been
presented in \cite{SS,LXCDM} with interesting implications on the
coincidence problem. The renormalization group equation (\ref{IR
CC}) leads to \ $\La=\La_0+\si M^2\,(H^2-H_0^2)/(32\pi^2)$.
Furthermore, there is the Friedmann equation $\,H^{2}=({8\pi G
}/{3})\left( \rho +\Lambda\right)\,$ and the conservation law,
which can be chosen in different ways. In one of the possibilities
\cite{CCfit} we can admit the energy exchange between the vacuum
and matter sectors (see also \cite{waga}), so that we have
$\dot\Lambda+\dot\rho+3\,H\,\rho=0$. The solution of this set of
coupled equations is completely analytical and the effect of the
running is parametrized by the dimensionless parameter \
$\nu\,=\,\sigma\,M^2/12\pi M_P^2\,.$ When $\nu\rightarrow 0$ we
recover the standard result for $\,\La=const$. The value of
$|\nu|$ has to satisfy the constraint $|\nu|\ll 1$, typically
$|\nu|\lesssim 10^{-2}$, for otherwise there is a dominance of DE
over radiation in the nucleosynthesis time \cite{CCfit}. It was
only later recognized that the strongest constraint actually comes
from the computation of the density perturbations \cite{CCwave},
where the consistency with the LSS galaxies distribution data
requires $|\nu|$ to be, at most, $10^{-4}$ with the best fit
corresponding to values smaller than $10^{-6}$. Qualitatively
similar results have been achieved earlier in the framework of
analogous models and, more quantitatively, in modified
quintessence models\,\cite{ana}.

The second RG framework \cite{Gruni} does not permit the energy
exchange between vacuum and matter sectors. This is a good point,
because {\it a)} the conservation law is nothing but a
mathematical expression of covariance. We have no reason to think
that the vacuum and matter effective actions are not separately
covariant; \ {\it b)} The energy exchange between vacuum and
matter assumes the creation of particles and the creation of both
massive and massless particles in the present-day universe meets
obvious obstacles. The conservation law for the vacuum action with
variable CC requires that the Newton constant becomes weakly
depending on the Hubble parameter. Investigation of density
perturbations in this model is an open problem and should be
explored soon.

\section{Conclusions}

\qquad We presented a short review of CC problems. The fine-tuning
of CC is a hierarchy problem due to the huge difference between
the particle physics and cosmological scales. Different approaches
for solving this problems have been developed. The anthropic
considerations, together with the idea of multiple vacuum states
coming from string theory, gives an important hint about possible
values of CC. The renormalization group, in the framework of
quantum field theory in curved space-time, indicates the
possibility of slowly varying CC even at the cosmic low energy
scale.
\vskip 2mm

{\bf Acknowledgments.}

Authors are grateful to R. Bousso and A. Vilenkin for useful 
correspondence concerning the anthropic approach. The work of 
I.Sh. has been partially supported by the research grants from 
CNPq (Brazil),  FAPEMIG (Minas Gerais, Brazil),  and by the 
fellowships from CNPq and ICTP(Italy). J.S. has been supported 
in part by MECYT and FEDER under project 2004-04582-C02-01, 
and also by DURSI
Generalitat de Catalunya under project 2005SGR00564


\end{document}